\documentclass[%
 reprint,
 amsmath,amssymb,
 xcolor, afterpage
pre,
]{revtex4-1}

\RequirePackage{bbm} 
\usepackage{mathtools}
\usepackage{graphicx}% Include figure files
\usepackage{dcolumn}% Align table columns on decimal point
\usepackage{bm}% bold math
\usepackage{xcolor}
\colorlet{new}{black}
\usepackage{natbib}
\usepackage{cleveref}

\crefname{equation}{equation}{Eq.}
\Crefname{equation}{Eq.}{Eq.}
\crefname{table}{table}{tables}
\Crefname{table}{Tab.}{Tab.}

\crefname{figure}{figure}{figures}
\Crefname{figure}{Fig.}{Fig.}

\begin{document}

\title{Constrained basin stability for studying transient phenomena in dynamical systems}

\author{Adrian van Kan$^{1,2,}$}
 \email{van\_kan@stud.uni-heidelberg.de}
 \email{Both corresponding authors contributed equally to the reported research.}
\author{Jannes Jegminat$^{1,}$}%
 \email{jegminat@iup.uni-heidelberg.de}
\affiliation{%
$^1$Department of Physics and Astronomy, University of Heidelberg, Im Neuenheimer Feld 226, D-69120 Heidelberg, Germany\\
$^2$Department of Physics, Imperial College London, Prince Consort Rd, London SW7 2BB, UK
}%

\author{Jonathan F. Donges$^{3,4}$, J\"urgen Kurths$^{3,5,6,7}$}
\affiliation{
 $^3$Potsdam Institute for Climate Impact Research, P.O. Box 601203, D-14412 Potsdam, Germany \\
 $^4$Stockholm Resilience Centre, Stockholm University, Kr\"aftriket 2B, 114 19 Stockholm, Sweden
}%
\affiliation{
 $^5$Department of Physics, Humboldt University Berlin, Newtonstr.~15, D-12489 Berlin, Germany
}%
\affiliation{
 $^6$Institute for Complex Systems and Mathematical Biology, University of Aberdeen, Aberdeen AB24 3FX, United Kingdom
}%
\affiliation{
 $^7$Department of Control Theory, Nizhny Novgorod State University, Gagarin Avenue 23, 606950 Nizhny Novgorod, Russia
}%

\date{\today}% It is always \today, today,

\begin{abstract}
% Background: Basin stability, early warning signal of bifurcations, transients
Transient dynamics are of large interest in many areas of science. 
% Methods: constrained basin. Condition used to discriminate transients. 
Here, a generalization of basin stability (BS) is presented: constrained basin stability (CBS) that is sensitive to various different types of transients arising from finite size perturbations. 
% Results: Early warning for explosive bifurcation. Information about transient dynamics in simple measure.
\textcolor{new}{CBS is applied to the paradigmatic Lorenz system for uncovering nonlinear precursory phenomena of a boundary crisis bifurcation. Further, CBS is used in a model of the Earth's carbon cycle as a return time-dependent stability measure of the system's global attractor.
% Conclusions: Intermediate measure between linear and basin stability. Combination of asymptotic and dynamical properties into one measure. 
Both case studies illustrate how CBS's sensitivity to transients complements BS in its function as an early warning signal and as a stability measure.}
% Relevance:
CBS is broadly applicable in systems where transients matter, from physics and engineering to sustainability science. Thus, CBS complements stability analysis with BS as well as classical linear stability analysis and will be a useful tool for many applications.

\begin{description}
\item[PACS numbers]
05.45.-a, 89.90.+n
\end{description}
\end{abstract}

\pacs{Valid PACS appear here}% PACS, the Physics and Astronomy
\maketitle

%\tableofcontents

\section{Introduction}
\label{Introduction}

Many fields of science analyze dissipative dynamical systems in terms of their attractors. Thus, it is an important challenge to quantify the stability of attractors with respect to a given perturbation. The most popular method is linear stability analysis which considers infinitesimal perturbations. \citet{Menck13} suggest to complement this linear measure with basin stability (BS) which accounts for finite and even large perturbations. The application of BS to power grids has yielded novel mitigation strategies against super outages \citep{menck2014dead,schultz2014detours}. BS is computed by estimating the volume of an attractor's basin. Therefore it is not sensitive to different forms of transient dynamics. However, transient phenomena in complex systems are of large interest in many areas of science, such as climatic and, more generally, global change in Earth system science~\cite{IPCC2013}, epileptic seizures in neuroscience~\cite{rabinovich2006dynamical}, ecosystem transitions in ecology~\cite{scheffer2009early} as well as in the previously mentioned study of super outages in power grids~\citep{menck2014dead,schultz2014detours}. For example, in the case of climate change~\cite{IPCC2013} and the great acceleration~\cite{steffen2015trajectory} as transient phenomena in the global social-environmental system~\cite{Schellnhuber1998,Schellnhuber1999}, major efforts are invested into studying the maximum global mean temperature and its timing along the trajectory due to anthropogenic greenhouse gas emissions. Moreover, the model- and data-driven analysis of transient global change trajectories underlies many recently proposed frameworks for sustainable development such as tolerable environment and development windows~\cite{petschel1999tolerable}, planetary boundaries~\cite{rockstrom2009,steffen2015planetary} and the safe and just operating space for humanity~\cite{Raworth2012}.\\
\textcolor{new}{ Making BS sensitive to transients, we generalize it to a family of stability measures termed constrained basin stabilities (CBSs). As opposed to BS, CBS is not computed from the entire basin of an attractor but only from a subset of the basin. The subset is defined by a generic constraint imposed on the transients. Thus, CBS is sensitive to transients while, maintaining the intuitiveness and simplicity of BS. To illustrate how CBS complements BS, we choose two specific constraints on transients, one based on the confinement of transient trajectories to certain regions in phase space and one based on transient duration, and apply them to the Lorenz system and a global carbon cycle model respectively. In the former example, CBS anticipates a boundary crisis bifurcation. In the latter, we show that CBS represents a more intuitive measure for stability than BS because CBS reflects not only that perturbation-induced transients return to the attractor but also that they do so within a desirable time interval.}\\
This paper is structured as follows. In Sec.~\ref{Methods} we introduce CBS and discuss some of its properties used in the further analysis. In Sec.~\ref{Application} we present two examples of CBS analysis in dynamical systems: the paradigmatic \citet{lorenz1963} model and a global carbon cycle model proposed by \citet{anderies2013topology}. Then, in Sec.~\ref{Discussion} we discuss the relevance of CBS and how it differs from established stability concepts. The paper concludes with closing remarks.

\section{Methods}
\label{Methods}

Let the \textcolor{new}{(not necessarily analytic)} vector valued function $f$ represent an autonomous potentially multistable dynamical system $\dot{\textbf{x}} = f(\textbf{x})$, where $\mathbf{x} \in \mathbb{R}^n$, and let $\phi^t(\textbf{x}_0)$ be the system state at time $t$ on a trajectory starting at $\textbf{x}_0$ at $t=0$. The basin stability BS($A$) of an attractor $A$ of this system quantifies the probability that after a finite size perturbation trajectories return to $A$~\cite{Menck13}. Perturbations within the attractor's basin $\mathcal{B}(A)$ return but the remaining ones fall into a different attractor. The probability distribution of perturbing a trajectory on the attractor to the state $\textbf{x}$ is given by $\rho(\textbf{x})$. Thereby, BS is formally defined as
\begin{align}
\label{eq:Basin-Stability}
\mathrm{BS}(A) = \int_\Gamma \mathrm{d}x^n \rho(\textbf{x}) \mathbf{1}_{\mathcal{B}(A)}(\textbf{x}),
\end{align}
where $\Gamma$ denotes the state or phase space of the dynamical system and the indicator function is
\begin{align}
\mathbf{1}_{\mathcal{B}(A)}(\textbf{x}) = 
\begin{cases} 
      \hfill 1   \hfill & \text{if  } \textbf{x} \in \mathcal{B}(A) \\
      \hfill 0 \hfill & \text{else.} \\
  \end{cases}
\end{align} 
BS can be computed quickly once an attractor's basin is given: it equals the mass of $\rho$ that is supported by the basin. However, in practice, the basin of attraction is usually not known and needs to be determined first. For this purpose, initial conditions are sampled according to the perturbation density $\rho$ and then integrated until they reach an attractor. Thus, in computing BS only the long-term limit of the trajectories is used to determine if an initial condition lies in the basin of attraction. Therefore, by construction, BS does not depend on transient motion.\\
To generalize BS, here we propose instead to use the properties of transients to define a class of stability measures that we term \textit{constrained basin stabilities} (CBSs). Calculating CBS requires a computational effort similar to that needed for BS, but CBS reveals additional information about the system that is encoded in the transient trajectories.
We define a transient as the set of points belonging to the part of the trajectory between the initial condition $\mathbf{x}(0)=\textbf{x}_0$ and reaching the attractor $A$,
\begin{equation}
T(\mathbf{x}_0) = \{ \phi^t(\mathbf{x}_0)\in \Gamma\backslash A\text{ } | \text{ } t\geq 0\}.
\end{equation}
\textcolor{new}{The fact that we define the attractor not to be part of the transient makes a difference for example in the case of trajectories induced by non-smooth flows where an attractor may be reached within finite time.} The idea of CBS is that a region in phase space is identified by some constraint on the transients starting from a subset of phase space
\begin{align}
C = \lbrace \mathbf{x}\in\Gamma \backslash A | \text{ the transient from } \mathbf{x} \text{ satisfies a constraint} \rbrace .
\end{align}
%\begin{align}
 %X(\textbf{x}_0) := \{ (t,\phi^t(\textbf{x}_0) )  \hspace{0.1cm} | \hspace{0.1cm}  t \in \mathbb{R}, \text{ }\phi^t(\textbf{x}_0) \in \Gamma / A\},
%\end{align}
%where $\phi^t(\mathbf{x}_0)$ is the flow of the dynamical system under consideration. 
%\begin{align}
%C(X(\mathbf{x}_0)) = 
%\begin{cases} 
 %     \hfill 1   \hfill & \text{if  } X(\mathbf{x}_0) \text{ fulfills } c \\
   %   \hfill 0 \hfill & \text{else.} \\
 % \end{cases}
%\end{align}
In other words, the transients starting from this \textit{conditioned set} $C$ satisfy the given constraint. For instance, we can choose $C$ to be the set of states $\mathbf{x}$ the transients starting from which exhibit monotonicity in the $x_1$ component. This is equivalent to demanding that the projection of a transient's velocity onto the basis vector $\textbf{e}_1$ in $x_1$-direction is non-vanishing. Thus, the conditioned set is $C_{mon} = \{\textbf{x} \in \Gamma \backslash A | f(\phi^t(\mathbf{x})) \cdot \textbf{e}_1 \neq 0 \, \forall \ t > 0\}$. If $x_1$ represents a population, the set $C_{mon}$ is the set of initial conditions which do not lead to a population overshoot~\cite{Brander1998}.\\ Incorporating an arbitrary constraint (not necessarily monotonicity) as an additional factor in \Cref{eq:Basin-Stability}, we formally define CBS as
\begin{align}
\label{eq:Cond_Basin-Stability}
\mathrm{BS}^C(A) = \int_\Gamma dx^n \rho(\textbf{x}) \mathbf{1}_C(\mathbf{x}) \mathbf{1}_{\mathcal{B}(A)}(\textbf{x}).
\end{align}

\begin{figure}[h!]
	\includegraphics[width=0.5\textwidth]{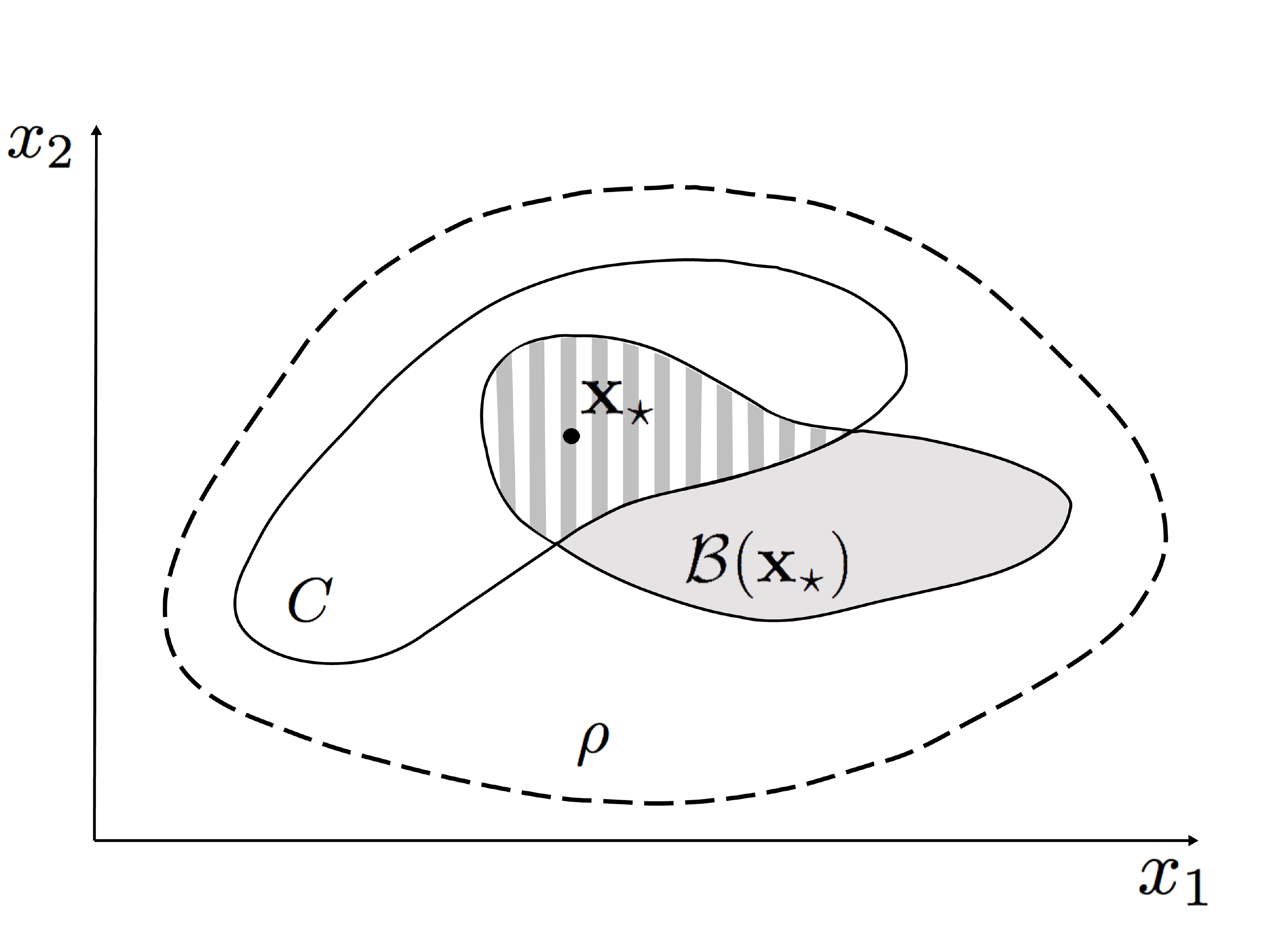}
	\caption{Subspaces of a schematic two-dimensional phase space containing a fixed point $\mathbf{x}_\star$: \textcolor{new}{perturbations are sampled from the domain of the perturbation probability density  $\rho$ (area within dashed line). Their transients return to the fixed point when sampled from the basin of attraction $\mathcal{B}(\textbf{x}_\star)$ (gray area). The set $C$ (white within solid line) is the set of initial conditions which lead to transients that satisfy a given constraint, e.g. on the $x_1$-component. While BS is computed as the fraction of perturbations within the basin of attraction, CBS is the fraction of perturbations within the intersection $C \cap \mathcal{B}(\textbf{x}_\star)$ (striped area). Thus, CBS reflects the stability with respect to perturbations whose transients fulfil a given constraint.}}
    \label{Sketch}
\end{figure}

The product of the two indicator functions checks whether a perturbation is inside of the attractor's basin and at the same time inside the region transients originating from which satisfy the prescribed constraint. \textcolor{new}{\Cref{Sketch} illustrates the regions in a schematic two-dimensional phase space that are relevant for computing BS and CBS for a fixed point.}
Three useful properties of CBS follow directly from its definition. Firstly, since $\mathbf{1}_C(\mathbf{x}) \leq 1$,
%\iffalse
\begin{align}
\mathrm{BS}(A) \geq \mathrm{BS}^C(A).
\end{align}
Secondly, let $\{C_i\}_{i\in I}$ be a partition of $\Gamma$, then
\begin{align}
\label{eq:Property2}
\sum_{i \in I}\mathrm{BS}^{C_i}(A) = \mathrm{BS}(A).
\end{align}
Thirdly, if $C_1 \subset C_2$ then
\begin{align}
\mathrm{BS}^{C_1}(A) \leq \mathrm{BS}^{C_2}(A).
\end{align}
The novelty of CBSs is that they integrate information about the transients into the asymptotic framework of BS. This information is encoded in a set $C$ of states the transients originating from which satisfy a given requirement, such as monotonicity in the previous example. We suggest classifying these requirements as static, dynamic and integrated, depending on how much information is necessary to find out which desirable region of phase space corresponds to them: (i)~Static requirements a-priori define a phase space region $\Gamma' \subset \Gamma$ that must not be entered by the transient. No further knowledge of the system or its dynamics is required. Examples are planetary boundaries in Earth system dynamics~\cite{rockstrom2009,steffen2015planetary}, minimum/maximum operating temperatures of a device or the evaluation of external functions (not $f$) of the system, e.g. the performance of a second system that depends on the system state $\textbf{x}$. (ii)~Dynamic requirements depend on velocity, thus more knowledge about the system is required: the dynamics, i.e. $f$, must be known. Using this knowledge, a region similar to $\Gamma'$ is defined. Consider, for example, a roller coaster that must not exceed a certain velocity or acceleration or the requirement of monotonicity in economic output to exclude the burst of market bubbles. (iii)~Integrated conditions depend not only on the current state of the transient but also on its past, i.e. they operate on an infinite dimensional space and memory effects are possible. Despite this complexity, testing integrated conditions is often easy in practice as can be seen from the following examples: imposing a limited number of opinion changes of a political party, thresholding the time needed to reach an attractor, integrated damage in a climate model or imposing a minimum average power output of a wind farm per time interval. Note that each of the constraints implies a binary decision: CBS identifies a qualitative property in a transient.
\medskip\\
In order to implement CBS (Eq.~\ref{eq:Cond_Basin-Stability}) numerically, we need to discretize it. For simplicity of presentation, we choose the attractor to be a fixed point, $A=\left\lbrace \mathbf{x}_\star \right\rbrace$ and consider a uniform distribution $\rho$ of $N$ initial conditions drawn from some subset of phase space approximated by \textcolor{new}{a set of sampling points} $\mathbf{x}_i$, $i\in \{1,\dots,N\}$  drawn at random from the phase space volume in question. This results in
\begin{equation}
\label{eq:cbs_discrete}
\mathrm{BS}^C(\mathbf{x}_\star, \varepsilon) = \frac{1}{N} \sum\limits_{i=1}^N \mathbf{1}_C(\mathbf{x}_i)\text{ } \Theta\left (\varepsilon-d_{min}\right ),
\end{equation}
\textcolor{new}{where $d_{min}$ is the minimal state-space distance (within the finite simulation time) between the fixed point $\mathbf{x}_\star$ and the transient $T(\mathbf{x}_i)$ and $\Theta(x)$ is the Heaviside function.} %defined as

If a trajectory reaches a distance smaller than the threshold $\varepsilon$ from the attractor within finite simulation time, we regard it to have reached the attractor. Furthermore, the uniformity of $\rho$ implies $\rho(\mathbf{x}_i) = N^{-1}$. Operationally, for any attractor $A$ (not necessarily a fixed point), we proceed as follows: 1. Sample an initial condition $\mathbf{x}_i$ according to $\rho$, 2. Integrate $\mathbf{x}_i$ in time until it has reached an attractor, 3. If the reached attractor is $A$, count $\mathbf{x}_i$ towards $\mathrm{BS}(A)$, 4. Check if the transient originating from $\mathbf{x}_i $ satisfies the constraint, if so count $\mathbf{x}_i$ towards $\mathrm{BS}^C(A)$, 5. Increase $i\to i+1$, repeat until $i=N$.\\
The computational procedure outlined above by which we determine BS and CBS allows us to estimate the uncertainty of our estimates of BS and CBS. Since we consider a uniform perturbation $\rho$, we are effectively drawing initial conditions at random from the subset $R$ of phase space where $\rho$ is non-zero. The fraction $p$ of the volume the basin $\mathcal{B}(A)$ occupied by $R$ is the true BS, i.e. the probability that we draw an initial condition from $\mathcal{B}(A)$ at random. This implies that effectively, our estimate of BS after drawing $N$ initial conditions comes from a binomial distribution with expectation value $p=\mathrm{BS}$, which leads to the standard deviation
\begin{equation}
\label{eq:sigma_BS}
\sigma_{\mathrm{BS}(A)} = \frac{1}{N}\sqrt{\mathrm{BS}(1-\mathrm{BS})N} = \frac{1}{ \sqrt{N}} \sqrt{\mathrm{BS}(1-\mathrm{BS})}.
\end{equation}
Equation (\ref{eq:sigma_BS}) also holds when $\mathrm{BS}(A)$ is replaced by $\mathrm{BS}^C(A)$, which follows from an argument analogous to the one above. It is important at this point to note that non-uniform distributions $\rho$ are also admissible and make sense in certain applications when some perturbations need to be weighted more than others. However, an error estimate as in \Cref{eq:sigma_BS} is less straightforward to obtain for non-uniform $\rho$.

\section{Application}
\label{Application}

To illustrate the versatility of CBS, we give examples of specific constraints in the paradigmatic Lorenz system \cite{lorenz1963} and in a global carbon cycle model by \citet{anderies2013topology}. In the Lorenz63 (L63) model we show how CBS can reveal precursory phenomena  before the onset of a boundary crisis bifurcation. In the Anderies model, we argue that CBS reflects our intuition of stability of a desired state against perturbations better than standard BS. We illustrate in both examples how CBS can generate important new insights into the dynamics of complex systems, while being simple enough to be amenable to a quick interpretation.	

\subsection{Anticipating a boundary crisis bifurcation}

\begin{figure}[h!]
\includegraphics[width=0.5\textwidth]{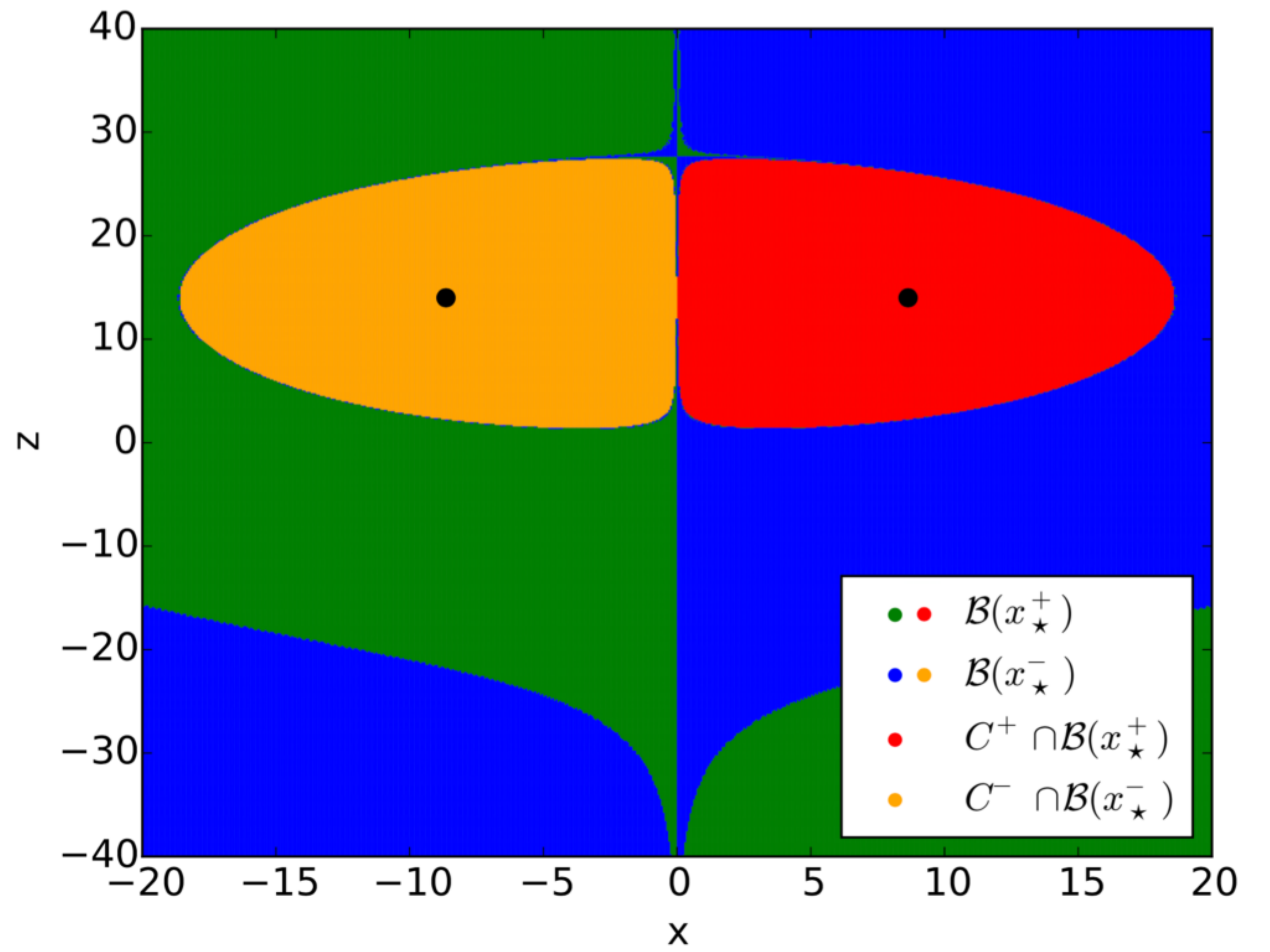}
\includegraphics[width=0.5\textwidth]{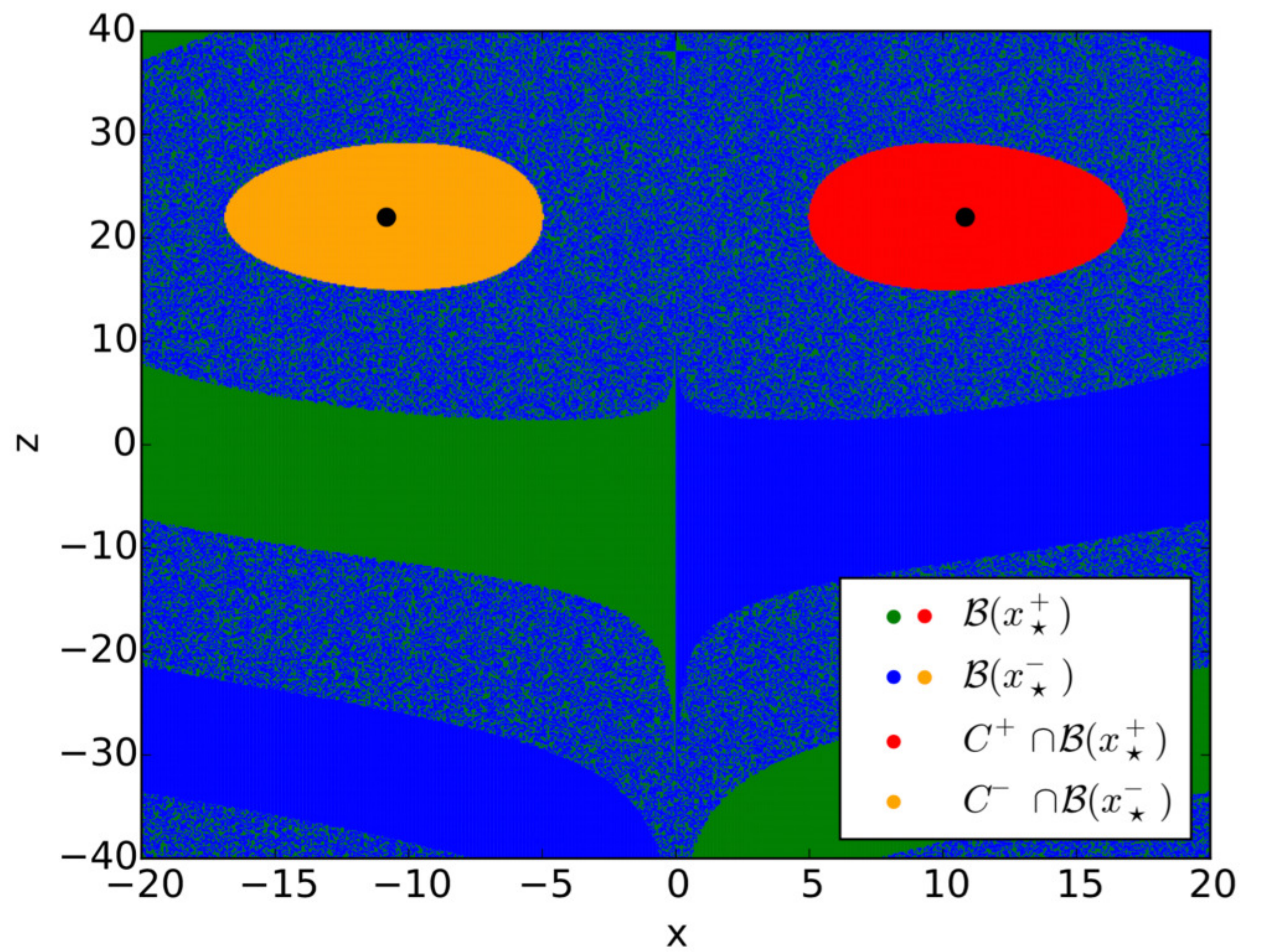}
\caption{(Color online) \textcolor{new}{Illustrations of the phase space structure of the L63 system defined by \Cref{eq:L63_i,eq:L63_m,eq:L63_f}. Both panels show cross sections, obtained by cutting along the plane containing the origin with normal $(1,1,0)$, of the basins of attraction $\mathcal{B}(\mathbf{x}_*^\pm)$ and their intersections $C^\pm \cap \mathcal{B}$ with the sets $C^\pm$ for $r=15$ (upper panel) and $r=23$ (lower panel). In both panels, the  blob-shaped region around each fixed point (black dots) is $C^\pm \cap \mathcal B$.
Top panel: the total basin of a fixed point is composed of successive layers: it is given by $C^\pm \cap \mathcal B$ combined both with the region in the respective other half-space ($x>0$ or $x<0$) directly surrounding $C^\pm \cap \mathcal B$ and with the next layer of the same color in the fixed point's half plane. 
In the lower panel (coloring identical), the fractal structure of the basins, visible as intermingled green and blue sets, is apparent; it is associated with transient chaos. One observes that the fraction of the window covered by the sets $C^\pm$ shrinks as $r$ increases. This illustrates the general behavior observed in the L63 system and quantified by CBS, namely that the volume fraction of the three-dimensional sampling region occupied by $C^\pm$ decreases continuously as $r$ approaches $r_3$.
}}
\label{CrossSectionL63}
\end{figure}

The L63-system \citep{lorenz1963} 
\begin{align}
\dot x &= \sigma (y - x) \label{eq:L63_i}\\
\dot y &= r x - y - xz \label{eq:L63_m}\\
\dot z &= xy - bz \label{eq:L63_f}
\end{align}
is a conceptual model of Rayleigh-B\'enard convection. It is famous for exhibiting chaotic dynamics along with a rich dynamical behavior. Setting $\sigma = 10$ and $b = 8/3$, we begin by summarizing the bifurcation structure as the parameter $r \in [9,26]$ increases. At first, two stable fixed points exist at $\textbf{x}_\star^{(\pm)} = (\pm \sqrt{b(r-1)},\pm \sqrt{b(r-1)},r-1)$, corresponding to left and right turning convection rolls respectively. At $r_1=13.926$ a chaotic saddle appears. At $r_2=24.06$ this chaotic saddle undergoes a boundary crisis and becomes attractive. The fixed points lose their stability at $r_3=24.74$. Our goal is to anticipate this boundary crisis \citep{strogatz2014nonlinear}. To this end, we choose a specific condition sensitive to long (chaotic) transients, since these are precursors of the crisis. With the following static constraint, we discriminate between transients that stay close to one of the fixed points ${\mathbf{x_+}} $ or ${\mathbf{x_-}} $ (i.e. one sense of convective overturning) and chaotic transients that flip between them:
\begin{align}
\label{eq:L63-condition}
C^{\pm} =  \{  \textbf{x} \in \Gamma \text{ } | \phi^t \text{ }   (\textbf{x})   \cdot    \textbf{n} \neq 0 \text{ } \forall t > 0 \}
\end{align}

where the difference vector $\mathbf{n} = |\mathbf{x}_\star^+-\mathbf{x}_\star^-|^{-1}(\mathbf{x}_\star^+-\mathbf{x}_\star^-)$ is the normal of a plane $H$ containing the origin that separates the phase space into two symmetric halves. \textcolor{new}{\Cref{CrossSectionL63} shows two-dimensional cross sections of the three-dimensional basins of attraction $\mathcal{B}(\mathbf{x}_*^\pm)$ and their intersections $C^\pm\cap \mathcal{B}(\mathbf{x}_*^\pm)$ with the sets $C^\pm$ defined above for two different values of the parameter $r$.}

\begin{figure}[h]
	\centering
	\includegraphics[width=0.53\textwidth]{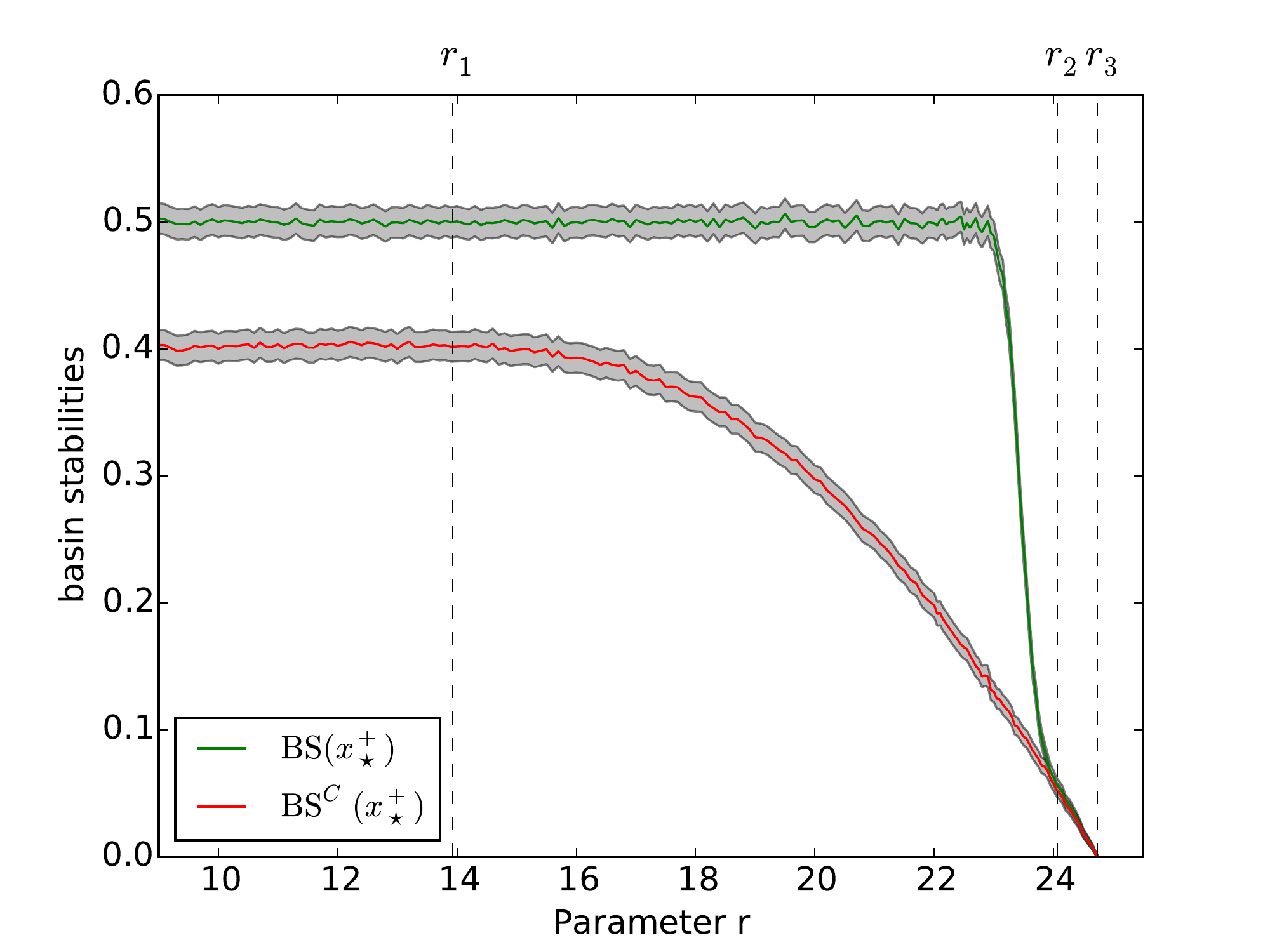}
	\caption{(Color online) \textcolor{new}{BS (green (upper) line) and CBS (red (lower) line)} of the positive fixed points in the L63 system as the parameter $r$ is varied. From left to right, the vertical lines indicate the appearance of the chaotic set ($r_1$), the boundary crisis ($r_2$) and the stability loss of the fixed point ($r_3$). The difference between the two curves represents the fraction of the basin from where chaotic transients evolve. Note that BS exhibits a jump at $r_2$ which is blurred by the finite simulation time: the simulation stopped before all (increasingly long) transients had reached the fixed point. Because of the system's x-y symmetry the negative fixed point has the same BS and CBS. The grey envelopes represent $\pm 3\sigma$ according to \Cref{eq:sigma_BS}. }
	\label{BS_L63}
\end{figure}

BS and CBS are computed for $15,625$  initial conditions sampled from the uniform perturbation distribution $\rho(\textbf{x}) = \tfrac{1}{40}\Theta(20 - |x|)\tfrac{1}{40}\Theta(20 - |y|)\tfrac{1}{30}\Theta(15-|z|)$, which describes a box that roughly covers the attractor. To compute BS, we evolve each initial condition in time until it reaches either one of the fixed points or the chaotic attractor. The termination condition in the former case is that the trajectory enters an $\varepsilon$-ball around $\mathbf{x}_\star ^{\pm}$, here $\varepsilon=10^{-4}$. In the latter case, we iterate until the trajectory has crossed the plane $H$ a large (but computationally feasible) number $m$ of times, here $m=400$.
\Cref{BS_L63} shows the resulting BS and CBS. The BS and CBS curves are identical for both fixed points due to their symmetry. Thus, only the values for the positive fixed point are shown.
For any conditioned set $C$ and its complement $\overline{C}=\Gamma \backslash C$, \Cref{eq:Property2} reduces to $\mathrm{BS}(A) = \mathrm{BS}^C(A) + \mathrm{BS}^{\overline{C}}(A)$. This implies that the difference between the two stability measures reflects the fraction of the basin from where (long) chaotic transients originate. The magnitude of CBS reflects the opposite, i.e. the part of the basin from where trajectories fall into the fixed point without crossing $H$. For $9 \leq r \leq 14$, the two stability measures are constant but differ in their value. For $14\leq r\leq 23$ the fraction of chaotic transients increases continuously (in agreement with \Cref{CrossSectionL63}), while the basin volume, i.e. BS, does not change. Between $r \approx 23$ and the bifurcation point at $r_2$, the basins of the fixed points collapse as the fraction of trajectories exceeding $m$ crossings grows rapidly. In the limit $m \rightarrow \infty$, the basin size changes discontinuously at $r_2$: the fraction of the basin that corresponds to the chaotic transients at $r<r_2$ suddenly feeds the newly born attractor when $r>r_2$. In \Cref{BS_L63} the drop is not discontinuous due to finite simulation time: long transients do not reach the fixed points before the simulation is ended.  From $r_2$ to $r_3$, both stability measures coincide because chaotic transients are absent due to the chaotic saddle now being attractive. For $r>r_3$, both BS and CBS vanish. 
In short, with BS alone we cannot anticipate the crisis at $r_2$. However, combining it with CBS, the emergence of the chaotic set can be observed by the increasing fraction of chaotic transients within the fixed points' basins. Although CBS does not predict the bifurcation at $r_2$, it does indicate the approaching crisis while BS does not and thus it substantially complements the original BS by revealing additional information on the basin structure.

\subsection{The Anderies carbon cycle model} % Model -> Supp Methods??
\begin{figure}[h!]
\includegraphics[width=0.515\textwidth]{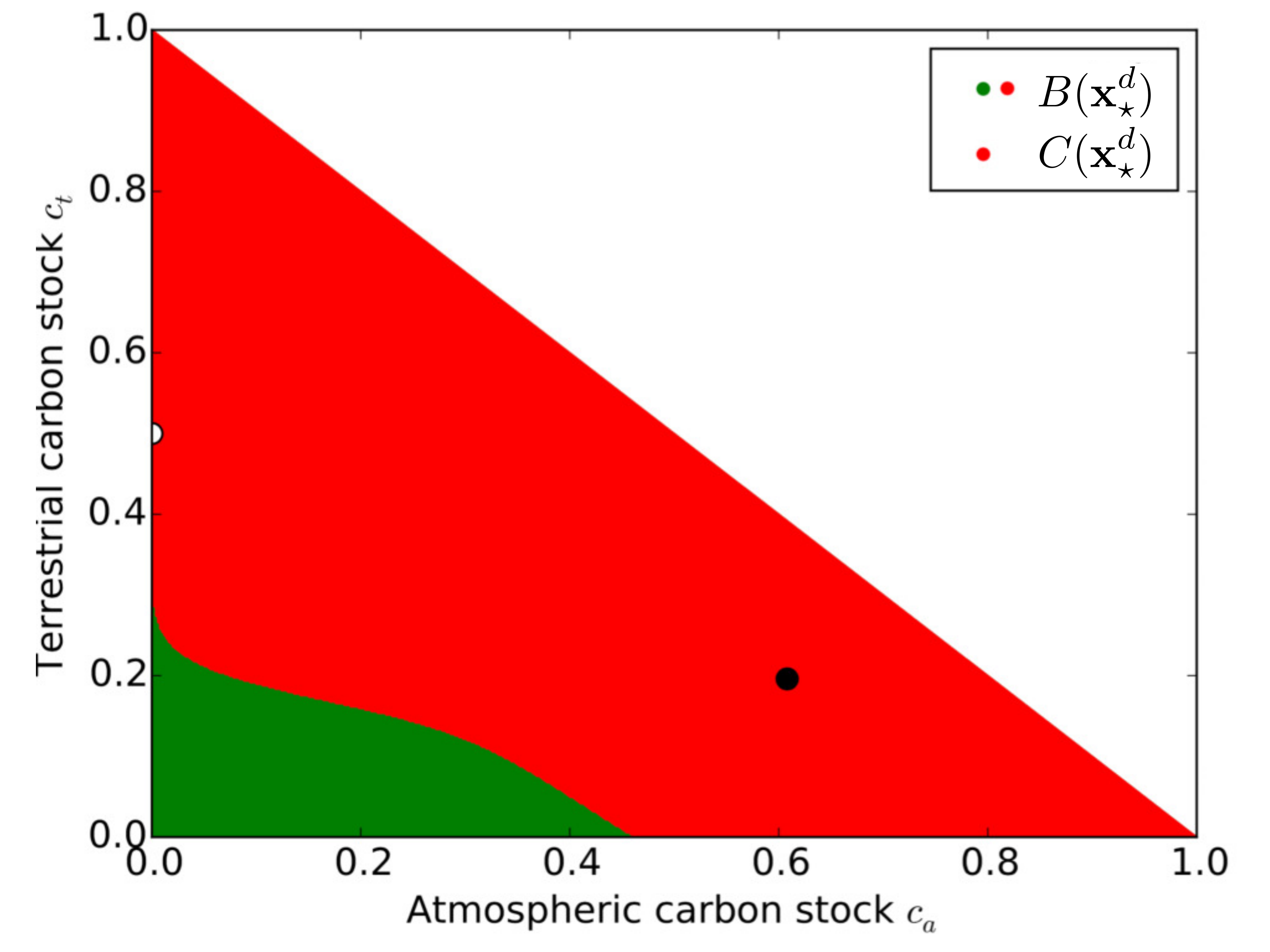}
\includegraphics[width=0.5\textwidth]{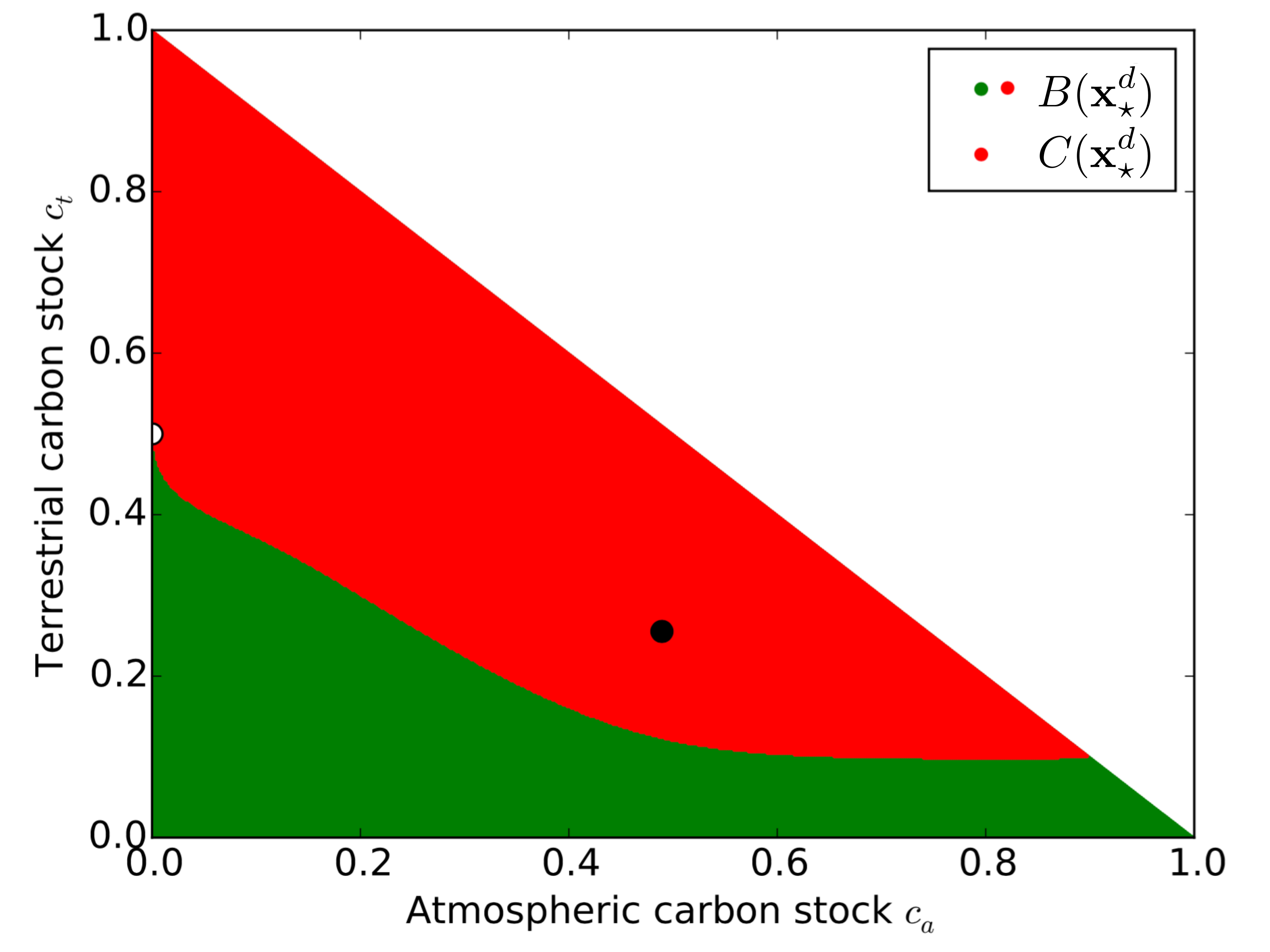}
\caption{(Color online) \textcolor{new}{Illustrations of the phase space structure of the Anderies system, \Cref{eq:Andi_i,eq:Andi_f}. Both panels show the (globally attracting) desirable fixed point (black dot) and the set $C$ (red, upper part of triangle) and its complement $\mathcal{B}/C$ (green, lower part of triangle) for $\alpha=0.15$ (upper panel) and $\alpha=0.35$ (lower panel). The white dot at $(0,0.5)$ is the desert state fixed point. One observes that the fraction of the two-dimensional finite phase space covered by the set $C$ shrinks as $\alpha$ increases, in agreement with \Cref{BS_CBS_Anderies}.}}
\label{cuts_anderies}
\end{figure}

\begin{figure}[h]
	\centering
	\includegraphics[width=0.53\textwidth]{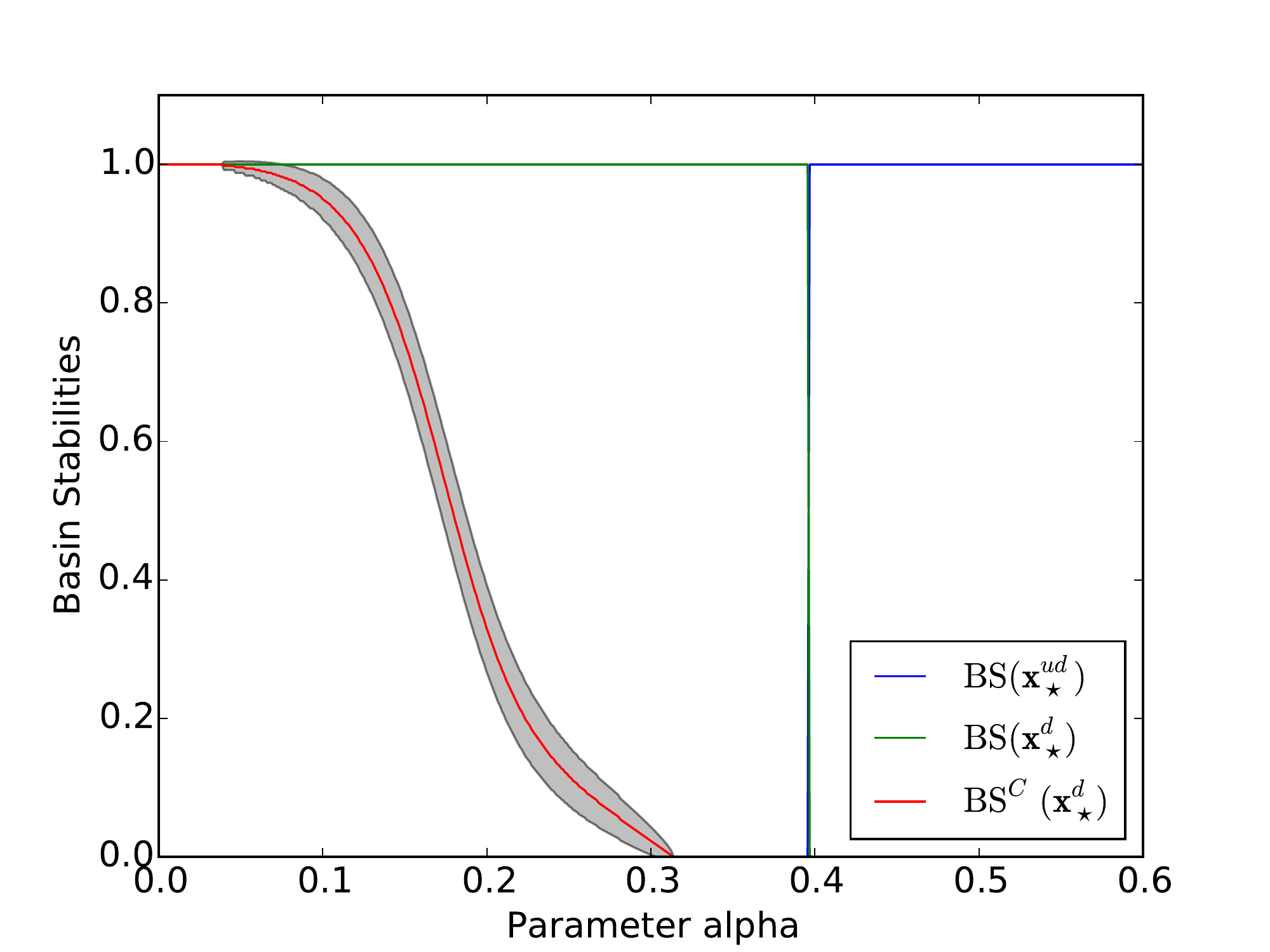}
	\caption{(Color online) \textcolor{new}{ BS (straight lines) and CBS (curved red line) of $\mathbf{x}_\star^d$ and $\mathbf{x}_\star^{ud}$ versus the human carbon offtake rate $\alpha\in[0,0.6]$ for $t_{crit} = 90$. BS of $\mathbf{x}_\star^d$ is represented by the  green straight line ($\alpha\leq \alpha_{crit}\approx 0.4$), BS of $\mathbf{x}_\star^{ud}$ by the blue straight line ($\alpha\leq \alpha_{crit}$).} At low values of $\alpha$, the desirable state is stable against any strength of perturbation while the desert state is unstable. For $\alpha > 0.03$, an increasing fraction of perturbation-induced trajectories take longer than $t_{crit}$ to return to the desirable fixed point until at $\alpha \approx 0.32$, CBS vanishes. By contrast, BS of both fixed points exhibits a jump at $\alpha_{crit}$ and  thus no precursory phenomena can be observed there. The grey envelope represents $\pm 3 \sigma$ according to \Cref{eq:sigma_BS}.}
	\label{BS_CBS_Anderies}
\end{figure}

\citet{anderies2013topology} present a conceptual model of global carbon cycle dynamics in the Earth system formulated as a mass balance between three carbon stocks $\mathbf{x}=(c_a,c_t,c_m)$ which are non-dimensional atmospheric $c_a$, terrestrial $c_t$ and marine stocks $c_m$ respectively. Formally, the model for the pre-industrial case is given by
\begin{align}
\dot{c_t} =& NEP(c_a,c_t)-H(c_t) \label{eq:Andi_i}\\
\dot{c_m} =& D(c_a,c_m), \label{eq:Andi_f}
\end{align}
where total carbon in the system is conserved such that $c_a+c_t+c_m = 1$ and $c_a,c_t,c_m\geq0$. The expressions describing the derivatives $\dot{c_t}, \dot{c_m}$ are defined as follows: a harvesting term $H(c_t)=\alpha c_t$, where $\alpha$ determines the human offtake of terrestrial carbon stocks, a diffusion term between atmosphere and ocean $D(c_a,c_m)=0.05(c_a-c_m)$ and net ecosystem productivity $NEP(c_a,c_t)=2.5 c_t \left( 1-c_t/0.7\right)  \lbrace 1.5 c_a^{0.3}  220 T(c_a)^3 \exp[-7T(c_a)]  -110 T(c_a)^4\exp[-5T(c_a)] \rbrace$ with $T(c_a) = 0.8c_a+0.2$.\\ It is found in \cite{anderies2013topology} that any initial condition converges to one of two fixed points of interest: either a desirable state $\mathbf{x}_\star^{d} = ((c_a)_\star^d,(c_t)_\star^d,(c_m)_\star^d)$ with vegetation or an undesirable global desert state $\mathbf{x}_\star^{ud}$.  At low values of $\alpha\in[0,0.6]$, the desirable state is attractive, while the undesirable state is repulsive. At $\alpha_{crit} \approx 0.4$ a transcritical bifurcation occurs and the fixed points reverse their stability. 
\citet{anderies2013topology} study their model in the context of planetary boundaries interacting with each other. In order to define a safe operating space, they suggest to classify trajectories by whether they return to a certain small $\varepsilon$-ball around the desirable fixed point $\mathbf{x}_\star^{d}$ by a certain critical time $t_{crit}$. Translating this into our framework, we obtain the condition
\begin{align}
\label{Andi-condition}
C =  \{  \textbf{x} \in \Gamma \text{ } | \text{ }  \exists t < t_{crit} \text{ s.t. } |\phi^t(\textbf{x}) -  \textbf{x}_\star^{d}  | < \varepsilon \},
\end{align}
\noindent where we choose $\varepsilon = 10^{-4}$. The constraint formulated in \Cref{Andi-condition} is an integrated constraint since it depends on time. The motivation for this choice of constraint is that, although all trajectories converge to $\mathbf{x}^{d}_\star$ as $t\to \infty$ for $\alpha<\alpha_{crit}$ (since then $\mathbf{x}^{d}_\star$ is globally attractive), some trajectories pass very closely and slowly by $\mathbf{x}_\star^{ud}$. These trajectories would entail catastrophic consequences for life on the planet. Therefore, they are identified by whether they exceed a certain return time threshold. Thus, we compute CBS of the desirable fixed point based on \Cref{Andi-condition}. We consider a perturbation density $\rho$ describing a depletion of the terrestrial carbon stock $c_t$ (e.g. by immense wildfires). The released carbon is fed into the atmospheric carbon stock $c_a$. We implement this scenario using a uniform perturbation density on a line in phase space: the terrestrial carbon stock is depleted to a value $c_t\in[0,(c_t)_\star^d)$ while the marine carbon stays constant, $c_m = (c_m)_\star^d$ and the atmospheric carbon increases according to the carbon conservation law $c_a = 1-c_m-c_t$. We draw $N=500$ initial conditions.  \Cref{cuts_anderies} shows the set $C$ in two-dimensional phase space for two different values of $\alpha$.
The more detailed dependence of BS and CBS on $\alpha$ is shown in \Cref{BS_CBS_Anderies}: BS is discontinuous at $\alpha_{crit}$ (within numerical accuracy), whereas CBS exhibits a smooth monotonic decay from 1 to 0 on the interval $\alpha \in[0.02,0.31]$. This reflects the fact that perturbations result in undesirable trajectories much more frequently as human carbon offtake increases until, at $\alpha\approx 0.31$, the return time for the considered perturbations always exceeds $t_{crit}$. Even though $\alpha<\alpha_{crit}$, none of the perturbed trajectories can avoid passing through a long quasi-global desert state. On this parameter interval the desired state is unstable with respect to CBS but stable with respect to BS. We suggest that the former measure provides a more meaningful notion of Earth system resilience from an anthropocentric point of view: it measures the probability that perturbations decay within a predefined acceptable time horizon, while the latter only measures the probability of returning within any (possibly infinite) time horizon. Further, CBS captures the change in transient structure and therefore reveals a signal of the transition in the Anderies model already at values of $\alpha$ significantly smaller than $\alpha_{crit}$. In contrast, BS is discontinuous (within numerical accuracy) at $\alpha_{crit}$ and does not exhibit any precursory phenomena. \textcolor{new}{\Cref{cuts_anderies} shows the set $C$ defined above in \Cref{Andi-condition} and $\mathcal{B}$, the basin of attraction of the desirable fixed point for two different values of $\alpha$.}

\section{Discussion}
\label{Discussion}

We have defined CBS as a generalization of BS, thereby combining an asymptotic stability measure with information retrieved from transient behavior into a compact and \textcolor{new}{intuitive measure. While BS is computed from an attractor's basin, CBS is computed from a subset of the attractor's basin. The subset is defined by the transient behavior of trajectories originating from this subset. Thus, CBS represents potentially very complicated transient behavior as an easy to interpret scalar quantity.\\
To underpin that a compact representation of transient behavior is highly relevant in applications, we have presented two examples using specific constraints on the transients.} In the case of Rayleigh-B\'enard \textcolor{new}{dynamics in the scope of the Lorenz63 model} we used the static constraint that the sense of rotation of convection rolls does not change. \textcolor{new}{Here, CBS uncovers nonlinear precursory phenomena of a boundary crisis bifurcation. In the global carbon cycle model by Anderies et al.~\cite{anderies2013topology}} we have studied the stability of the desirable state for a specific perturbation scenario under the premise that it can be restored within an acceptable time horizon. \textcolor{new}{CBS reflects the fact that long return times to the attractor after a perturbation are not desirable. More generally,} these applications demonstrate the three main advantages of CBS over BS: (i)~CBS provides useful information in the case of global attractors, \textcolor{new}{while BS cannot be meaningfully applied (it is always equal to 1).} (ii)~Sudden changes in basin size are often preceded by a change in transient behavior. Extending linear notions of early warning signals for incipient bifurcations~\cite{scheffer2009early}, CBS uncovers these nonlinear precursory phenomena in the case of the Lorenz63 model and helps anticipating the boundary crisis. (iii)~CBS reflects the fact that certain perturbation-induced transients are often undesirable, e.g. long return times\textcolor{new}{, thus allowing one to define highly relevant stability measures for a specific application.\\
The importance of BS lies in its applicability to a wide range of dynamical systems in various fields. The concept of CBS is even more general as it encompasses BS as special case. However, to apply CBS, we must choose a specific constraint, such as a limit on the return time. This choice strongly depends on a specific application, revealing highly relevant information there but potentially not being as useful in other applications. By providing two examples of useful constraints and by defining CBS precisely, formally and in close analogy to BS, we hope to facilitate the transfer of ideas between different applications and different generalizations of BS. For example, BS has been employed successfully to study power grid stability \cite{menck2014dead}. CBS could be used to develop more specific notions of stability, e.g. to impose that certain units recover quickly from mega outages or to constrain the total energy loss on the way of recovery. Another example is ecology where BS has proven to be a useful concept and transients are important \cite{dai2015relation}. CBS could be used to quantify questions of how fast ecosystems recover or investigate potential early warning signals based on minimal abundances of certain species after transients. More generally, BS has successfully applied in resilience research \cite{holling1973resilience,folke2004regime} and we expect interesting results from further investigating the notion of constrained resilience based on constraints on transients. We expect that future work on CBS will yield a set of transient-constraints that prove valuable across a wide range of different applications.}\\
CBS can be used in both passive and active experimental settings. In the former, we have only limited or no control of the system, e.g. the Earth system. We start with some normative notion of undesirable transients as the time threshold in the Anderies example. \citet{heitzig2015topology} discuss desirability in relation to phase space topology. From a given notion of desirability, a constraint is derived. Then, CBS addresses the question of how stable the system is with respect to perturbation-induced transients given that only some of them are desirable. If a system parameter varies over time, CBS is capable of revealing a stability trend which can justify an action to reverse the parameter change. It remains an open problem how CBS can be inferred experimentally or from observational data if a satisfactory model of the system is not available. In principle, if long time series of some environmental parameter (e.g. forest cover on the Earth's surface) can be derived from measurements and if many natural perturbations can be observed in the data, such as volcanic eruptions, then these can be exploited to estimate CBS. In the active setting, we use CBS to foster our understanding of the system without needing a normative proposition. The constraint and the perturbation are chosen such that new information about the structure of the basin of attraction is revealed. This situation is analogous to the Lorenz63-example: restricting the number of flips between the two halves of phase space (i.e. the two convection senses), the basin of attraction can be subdivided according to the number of flips. Thus, CBS helps to characterize a system that is subject to perturbations. In this active setting, it is easier to measure CBS: the system parameters can be chosen freely and the number of perturbations is not restricted by historic events. \\
The specific condition (\ref{eq:L63-condition}) is reminiscent partly of the concept of 'viability' \cite{aubin2007introduction,aubin2011viability}, although there are significant differences: in particular, in our case, there are no management options and more importantly, we are considering deterministic dynamics while viability theory incorporates stochastic and more generally non-deterministic processes. Furthermore, (\ref{eq:L63-condition}) can easily be generalized by allowing for trajectories to 'pierce through' $H$ once or multiple times - these generalizations are not related to viability theory. Another concept that has certain features in common with the condition (\ref{eq:L63-condition}) is 'survivability' \citep{hellmann2015survivability}. There too, a desirable region of phase space is designated as in the case of our choice of $C^{\pm}$. However, survivability does not incorporate the asymptotic nature of BS: it depends on the fraction of trajectories starting in a designated region of phase space spending a certain time exclusively in that region. In particular, it does not depend on which attractor trajectories converge to in the long-time limit. For these reasons, CBS is different from both viability theory and survivability and presents a novel and broadly applicable concept for quantifying stability of an attractor with respect to a given not only small perturbation, uniting both the asymptotic features of BS and the transient features of survivability. 
In conclusion, CBS represents a general framework to quantify the stability of attractors with broad applicability in various fields with an interest in complex dynamical systems, ranging from physics and technology to sustainability science.

\begin{acknowledgements}
The research reported in this article was developed within the working group ``Complex systems perspectives on anthropogenic climate change'' at the Natur- and Ingenieurwissenschaftliches Kolleg sponsored by the German National Academic Foundation (Studienstiftung des deutschen Volkes) and in the scope of the COPAN project on coevolutionary pathways at the Potsdam Institute for Climate Impact Research. JFD thanks the Stordalen Foundation (via the Planetary Boundary Research Network PB.net) and the Earth League's EarthDoc program for financial support. We thank four anonymous referees for their useful comments which helped to improve the readability of this paper and we are grateful for insightful discussions with Jobst Heitzig and Boyan Beronov. 
\end{acknowledgements}

\bibliography{newbib.bib}

\end{document}